\documentclass[twocolumn,showpacs,preprintnumbers,amsmath,amssymb]{revtex4}

\usepackage{graphicx}
\usepackage{dcolumn}
\usepackage{bm}

\begin{document}

\preprint{APS/123-QED}

\title{Trapping of ultra-cold atoms with the magnetic field of vortices in a thin film superconducting micro-structure}

\author{T. M{\"u}ller}\altaffiliation{Centre for Quantum Technologies, National University of Singapore, 3 Science Drive 2, Singapore 117543}
\author{B. Zhang}
\author{R. Fermani}\altaffiliation{Centre for Quantum Technologies, National University of Singapore, 3 Science Drive 2, Singapore 117543}
\author{K.S. Chan}
\author{Z.W. Wang}
\author{C.B. Zhang}
\author{M.J. Lim}\altaffiliation[Permanent address: ]{Department of Physics and Astronomy, Rowan University, 201 Mullica Hill Road, Glassboro, New Jersey, USA}
\author{R. Dumke}\email{rdumke@ntu.edu.sg}
\affiliation{Nanyang Technological University, Division of Physics and Applied Physics, 21 Nanyang Link, Singapore 637371
}

\date{\today}

\begin{abstract}
We store and control ultra-cold atoms in a new type of trap using magnetic fields of vortices in a high temperature superconducting micro-structure. This is the first time ultra-cold atoms have been trapped in the field of magnetic flux quanta. We generate the attractive trapping potential for the atoms by combining the magnetic field of a superconductor in the remanent state with external homogeneous magnetic fields. We show the control of crucial atom trap characteristics such as an efficient intrinsic loading mechanism, spatial positioning of the trapped atoms and the vortex density in the superconductor. The measured trap characteristics are in good agreement with our numerical simulations.\end{abstract}


\maketitle

Atom-optical systems combined with well-established superconductor technology allows a new generation of fundamental experiments and applications, potentially enabling a coherent interface between neutral atoms and solid-state quantum devices. Important applications include the quantum state transfer and manipulation between atomic and solid-state systems which is of great interest for quantum information. For this goal the combination of atomic or molecular quantum systems with quantum states in superconducting solid-state devices has been proposed in various forms \cite{Tian04,Sorensen04,Andre06,Rabl06,Petrosyan08,Tordrup08,Imamoglu09,Verdu09}. Recently, superconducting current-carrying chips have been used to implement micro-traps for neutral atoms \cite{Nirrengarten06,Mukai07,Cano08} and advantages over conventional chips have been shown \cite{Emmert09,Hufnagel09,Kasch09}. A prominent approach for quantum state manipulation in superconductors utilizes the magnetic flux quantum \cite{Makhlin01,Mooij99,Friedman00,Plantenberg07}. The flux quantum is of particular interest as an interface between atomic quantum systems and solid-state quantum devices because atoms with a magnetic dipole moment can be manipulated to high precision using magnetic fields. The pairing of atoms with quantized magnetic flux is a promising way of achieving a controlled interaction with possible applications in quantum technology and fundamental studies.\\
In this article we report the trapping of ultra-cold atoms that relies on the controlled coupling between vortices in a superconductor and the magnetic dipole moment of $^{87}\mathrm{Rb}$ atoms. This mechanism allows the design of novel trapping or guiding architectures for ultra-cold atoms. Such architectures could be additionally tailored by using combinations of vortices with magnetic fields induced by applied currents in superconducting micro-structures. This might be used to investigate the vortex dynamics in current-carrying superconducting micro-traps \cite{Scheel07,Nogues09}.\\
\begin{figure}
\includegraphics[width=8.6cm]{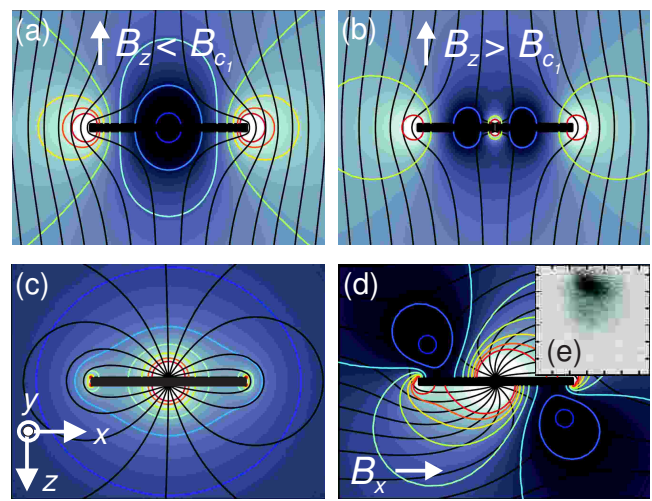}
\caption{\label{fig:1} (Color online) Schematic illustration of the magnetic field distribution during the preparation of the remanent-state superconductor (a)-(c) and the vortex-based micro-trap (d). Illustrated are the field lines (black lines) and equipotentials (colored lines and grayscale) of the magnetic field. An absorption image of $6\cdot 10^5$ atoms recorded 1ms after release from the vortex-based micro-trap is shown in (e).}
\end{figure}
The basic principles of the vortex-based trap are schematically shown in Fig.~\ref{fig:1}. The trap is created by a type-II superconducting thin film strip on a chip substrate. No magnetic field is applied as the superconductor crosses the transition temperature $T_c$. Below $T_c$ the film is ideally in the Meissner state, repelling any subsequently applied magnetic field smaller than the first critical field $B_{c_1}$ (Fig.~\ref{fig:1}(a)). To introduce vortices a magnetic field $B_{z,t_0}>B_{c_1}$ perpendicular to the chip surface is applied at time $t_0$ and field lines start penetrating the film (Fig.~\ref{fig:1}(b)). The magnetic field is raised to a value $B_{z,t_0}$ well above $B_{c_1}$ but below the second critical field $B_{c_2}$. Afterward the magnetic field $B_{z,t_0}$ is turned off. Due to the properties of the thin film, a large fraction of the vortices remains trapped. The vortices remaining in the film
are not isotropically distributed and the superconductor is referred as being in the remanent state \cite{Schuster94,Brandt96}. Figure~\ref{fig:1}(c) represents a simplified picture of the magnetic field due to a vortex trapped in the center of the strip. After preparing the trapped flux, the inhomogeneous field created by the vortices is combined with a magnetic bias field $B_{x,t_1}$ at time $t_1$ parallel to the surface of the thin film. This generates a field minimum below the strip (Fig.~\ref{fig:1}(d)), which is used to trap ultra-cold atoms.\\
We simulate the magnetic potentials of the vortex-based trap by means of mesoscopic models for magnetic flux penetration in type-II superconductors \cite{Schuster94,Brandt96,Dikovsky09}. In our simulations, we assume that for an increasing external field $B_{z}$ above $ B_{c_1}$, the vortices penetrate the superconductor from the edges and move toward the center of the strip. The spatial extent of the flux incursion $b$ from the edges depends on the applied field $B_z$, the critical current density $j_c$, the width $a$ and thickness $d$ of the strip according to the formula $b = \frac{a}{2}\left(1- \frac{1}{\cosh (\pi B_z /\mu_0 j_c d)}\right)$. The central region of the strip of width $\tilde{b}=a-2b$ remains flux-free \cite{Schuster94,Brandt96}. After the external field $B_z$ has been removed, our model considers the length scale $b$ as the fundamental parameter that characterizes the spatial distribution of the trapped magnetic flux. As the strip thickness $d$ is much smaller than its width $a$ and the typical trap-to-surface distance realized in our experiments, we neglect any effects of the magnetic field $B_{x,t_1}$ on the superconducting strip. Simulations of magnetic fields derived from the described model are shown in Fig. 2(b)-(c).\\
To realize the vortex-based magnetic trap we prepare a cloud of cold $^{87}$Rb atoms in an ultra-high vacuum chamber using standard laser
cooling and trapping techniques. Then the cloud is magnetically transported to the superconducting chip, which is mounted facing downward and cooled using liquid nitrogen. We cool the chip to about 83K without any additional thermal shielding.\\
The superconducting chip is a structured, $d=800\mathrm{nm}$ thin film of $\mathrm{YBa_2Cu_3O_{7-x}}$ (YBCO) on a yttria-stabilized zirconia single crystal substrate. The thin film has a critical temperature $T_c=89$K and at liquid nitrogen temperature $j_c=2.7\mathrm{MA/cm}^2$. We use a $400\mu\mathrm{m}$ wide strip to generate the main trapping potential. Using standard lithographic techniques the strip is structured in a Z-shape \cite{Denschlag99} which is commonly used in current carrying micro-traps. The central
strip has a length of 5mm and is crossed in its center by an additional wire of $200\mu\mathrm{m}$ width intended for trap compression \cite{Reichel01}. We use a bare YBCO film as additional metal layers can become the dominant source of magnetic near-field noise \cite{Emmert09,Kasch09}.\\
\begin{figure}
\includegraphics[width=8.6cm]{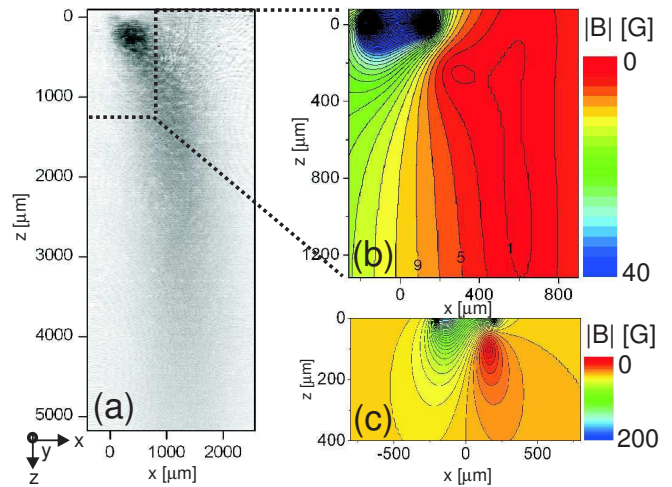}
\caption{\label{fig:2} (Color online) The superconducting strip is oriented along $y$ and its position is centered at $x=0$, $z=0$. (a) Absorption image showing the atomic density distribution in $trap_{M2}$. (b) Simulated magnetic field of $trap_{M2}$ in the region near the strip. (c) Simulated magnetic field of the trapped vortices combined with the bias field $B_{x,t_1}$. The field minimum below the edge of the strip corresponds to the radial trapping potential of $trap_{M3}$.}
\end{figure}
To load the atoms into the micro-trap and simultaneously prepare the superconductor in the remanent state, we employ the following
experimental sequence. A standard six-beam magneto-optical trap (MOT) is formed with typically $3\times 10^7$ atoms at a position 35mm below the chip center. After a compression MOT, molasses cooling and optical pumping to the $|F=2,m_F=2\rangle$ state we transfer the atoms to a magnetic trap ($trap_{M1}$), which is created by the same coils as used for the MOT. In this quadrupole trap the vertical magnetic field gradient $\frac{\partial B_{z,t_0}}{\partial z}$ can be varied between 20-35G/cm. Therefore, the magnetic field generating $trap_{M1}$ can efficiently be used to prepare the superconductor in the remanent state. At the location of the chip, the resulting field component $B_{z,t_0}$ perpendicular to the chip surface is 69-121G. This is well above the first critical field of our chip of about 25G (at 83K). After storing the atoms in $trap_{M1}$ for 160ms, they are adiabatically transferred to a second quadrupole-type magnetic trap ($trap_{M2}$), generated by a coil pair centered close to the chip surface. To transfer the atoms, we slowly ramp down the current for $trap_{M1}$ and simultaneously ramp up the current for $trap_{M2}$ in 460ms. The magnetic field component $B_{z}$ at the chip position is
reduced well below $B_{c_1}$ during the transfer. The vortices created by the magnetic field of $trap_{M1}$ remain in the thin film. We typically transfer $1.5\times 10^7$ atoms with a temperature of $120\mu\mathrm{K}$ to $trap_{M2}$.\\
The magnetic quadrupole-type field generated by the coils for $trap_{M2}$ is as expected significantly altered by the trapped vortices. In Fig.~\ref{fig:2}(a) we show the trapped atomic density distribution probed by \textit{in situ} absorption imaging along the strip axis $y$. The field minimum positioned 800$\mu\mathrm{m}$ below and 600$\mu\mathrm{m}$ to the side of the strip is due to the quadrupole field. Due to the combination of the vortex field and the fringing quadrupole field a second local minimum is formed below the edge of the strip. This is clearly seen in the high atomic density. This second minimum allows efficient transfer of the atoms to the vortex-based micro-trap ($trap_{M3}$). The observed atomic density distribution and the simulations of the total magnetic field shown in Fig.~\ref{fig:2}(b) are in good agreement.\\
The atoms are loaded into the vortex-based micro-trap $trap_{M3}$ by turning off the quadrupole field and applying a homogeneous bias field $B_{x,t_1}$. Strong radial confinement is realized along $x$ and $z$ by the combination of the vortices and the field $B_{x,t_1}$ (Fig.~\ref{fig:2}(c)). The weaker axial confinement along $y$ is provided by magnetic potentials resulting from vortices at the corners of the central strip, reminiscent of the Z-shape trap geometry \cite{Denschlag99}. In Fig.~\ref{fig:2}(c) we show the simulated contour lines of the total magnetic field in the $x-z$ plane which creates the radial trapping potential below the edge of the strip.\\
\begin{figure}
\includegraphics[width=7cm]{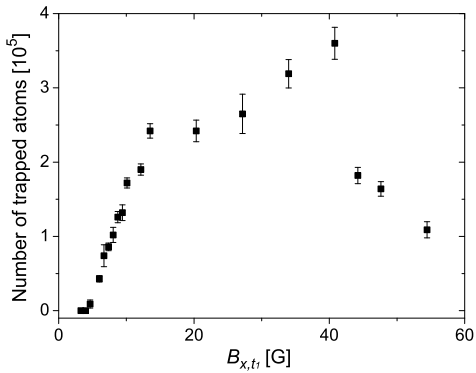}
\caption{\label{fig:3} Number of trapped atoms in the micro-trap $trap_{M3}$ for varying bias field $B_{x,t_1}$. A magnetic field $B_{z,t_0}$ of 112G has been used to prepare the vortices in the superconducting film for this measurement.}
\end{figure}
From $trap_{M2}$ we transfer up to $1\times 10^6$ atoms to $trap_{M3}$. Typically, we apply additional homogeneous bias fields $B_{z,t_1}$ and $B_{y,t_1}$ at time $t_1$ along the two other dimensions to improve the performance of the micro-trap. The field $B_{z,t_1}$ along the vertical direction shifts the atomic cloud position sideways along the width of the strip. This places the trap at different vortex densities as the flux is inhomogeneously distributed across the strip width \cite{Schuster94,Brandt96}. To decrease Majorana spin-flip losses, the additional bias field $B_{y,t_1}$ along the central strip is used to increase the absolute value of the magnetic field at the trap center. Typical bias fields for our trap are $B_{x,t_1}=40\mathrm{G}$, $B_{z,t_1}=7.7\mathrm{G}$ and $B_{y,t_1}=3.3\mathrm{G}$. With these parameters the trap lifetime is a few seconds, equal to the lifetimes of $trap_{M1}$ and $trap_{M2}$, which are limited by the background gas pressure in our single chamber vacuum setup.\\
To characterize the micro-trap we measure the atom number and trap-to-surface distance for various fields $B_{x,t_1}$ as shown in Fig.~\ref{fig:3} and Fig.~\ref{fig:4}. The atom number is inferred from absorption imaging after 1ms time of flight. With increasing $B_{x,t_1}$ the position of the trap center is shifted closer to the chip surface and therefore the radial magnetic field gradient at the trap center increases. The minimum field required to trap atoms as shown in Fig.~\ref{fig:3} is 4.6G. This corresponds to the magnetic field gradient at the trap center being just sufficient to overcome gravity. With increasing bias fields the depth of $trap_{M3}$ also increases. Therefore, we observe a sharp increase in the number of trapped atoms for bias fields from 4.6G to about 10G as a consequence from the atomic temperature in $trap_{M2}$. With further increasing $B_{x,t_1}$ the atom number starts to saturate, as the increase in trap depth and field gradient is countered by the simultaneous reduction of the trap volume of the micro-trap. For $B_{x,t_1}>41\mathrm{G}$ the size of the trapped cloud exceeds the distance to the chip surface and we observe loss of atoms.\\
\begin{figure}
\includegraphics[width=7cm]{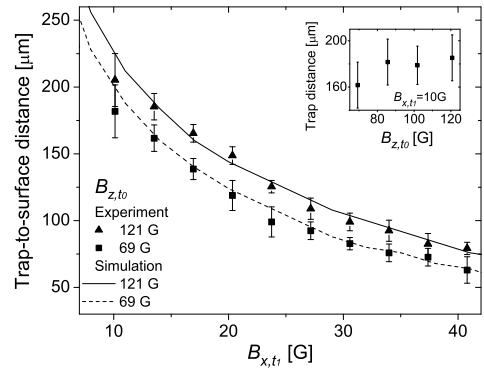}
\caption{\label{fig:4} Distance of the micro-trap $trap_{M3}$ to the chip surface for varying bias field $B_{x,t_1}$. The graph shows two series of distance measurements, each using a different magnetic field $B_{z,t_0}$ to create the trapped vortices. In the inset we show the measured trap distance for a fixed $B_{x,t_1}$ when $B_{z,t_0}$ is varied.}
\end{figure}
The trap-to-surface distance of $trap_{M3}$ is determined by \textit{in situ} imaging of the trapped atoms using a detection laser beam reflected from the chip surface \cite{Esteve04}. In Fig.~\ref{fig:4} we show the measured trap distance as a function of $B_{x,t_1}$. The distance is reduced for increasing $B_{x,t_1}$. However, for higher fields $B_{x,t_1}$ this effect diminishes as the trap distance becomes smaller than the spatial extent of the flux front $b$ on the strip.\\
We investigate the control of the amount of trapped vortices using the distance measurement displayed in Fig.~\ref{fig:4}. As the trap distance is determined by the cancellation of the average vortex magnetic field along $x$ by the bias field $B_{x,t_1}$ this measurement is a sensitive probe of the amount of trapped magnetic flux. We perform several series of distance measurements. For each series we use a different field $B_{z,t_0}$ to prepare the trapped vortices. Throughout all measurements the thin film is kept in the superconducting state and we increase $B_{z,t_0}$ from series to series. We did not observe any reduction of the amount of trapped flux during the measurements and the results agree well with the simulated trap-to-surface distance.\\
An estimate of the vortex density in the superconductor is $3-5\mu \mathrm{m}^{-2}$ for the range of $B_{z,t_0}$ used in our experiments. As a consequence, the average distance between individual vortices is still much smaller than the trap distance. Therefore, we do not resolve individual vortices and the atoms sample the field produced by the vortex density. However, by using atomic samples with sub-$\mu\mathrm{K}$ temperatures the magnetic potential in the near field of surface micro-traps can be resolved with high
precision \cite{Wildermuth05}. For such measurements, the vortex density could be varied with the externally applied magnetic field or by changing the chip temperature, e.g. by cooling with liquid helium.\\
A variety of fascinating experiments combining ultra-cold atoms and vortices in superconductors is within reach. Ultra-cold atoms
could be trapped by the combination of a single vortex and externally applied magnetic fields. This could be extended to time-varying potentials \cite{Petrich95,Colombe04}, to create novel types of magnetic traps or ring guides very close to a surface, e.g. below
$1\mu \mathrm{m}$. Periodic or quasi-periodic magnetic potentials created by vortex lattices could be investigated or used to manipulate the atoms. Moreover, ultra-cold atoms could be used for a spectroscopic measurement of the magnetic flux quantum.\\
In conclusion, we have realized the trapping of ultra-cold atoms with the magnetic fields of vortices in a remanent-state superconductor. We have shown the experimental control of important atom trap characteristics such as efficient loading, atom trap positioning and the vortex density in the superconductor, which agrees well with numerical simulations. The realized trap can be understood as a controlled interaction of atomic and solid-state quantum systems, paving the way for future discoveries in quantum technology and fundamental physics.
\begin{acknowledgments}
We thank A. Mohan, X. Wu and R.C. Ma for technical assistance regarding the experimental setup. We acknowledge financial support from Nanyang Technological University (grant no. WBS M58110036), A-Star (grant no. SERC 072 101 0035 and WBS R-144-000-189-305) and the Centre
for Quantum Technologies, Singapore. MJL acknowledges travel support from NSF PHY-0613659 and the Rowan University NSFG program.
\end{acknowledgments}

\end{document}